\documentclass[preprint]{vgtc}               

\graphicspath{{figures/}{pictures/}{images/}{./}} 

\usepackage{times}                     

\usepackage{tabu}                      
\usepackage{booktabs}                  
\usepackage{lipsum}                    
\usepackage{mwe}                       
\usepackage{subcaption} 
\usepackage{colortbl}
\usepackage{array}
\usepackage{arydshln} 
\newcommand{\subfigsCaption}[1]{\vspace{-.9em}\caption{#1}}

\usepackage{mathptmx}                  

\onlineid{1241}

\vgtccategory{System or Tool}

\vgtcinsertpkg

\title{reVISit-XR: Bringing Extended Reality into Embeddable, Trackable, and Replayable Visualization Studies}

\author{Shano Liang\thanks{e-mail: sliang1@wpi.edu} %
\and Max Chen\thanks{e-mail: mjchen@wpi.edu} %
\and Lane Harrison\thanks{e-mail: ltharrison@wpi.edu}}
\affiliation{\scriptsize Worcester Polytechnic Institute}

\teaser{
  \centering
  \includegraphics[width=0.83\linewidth]{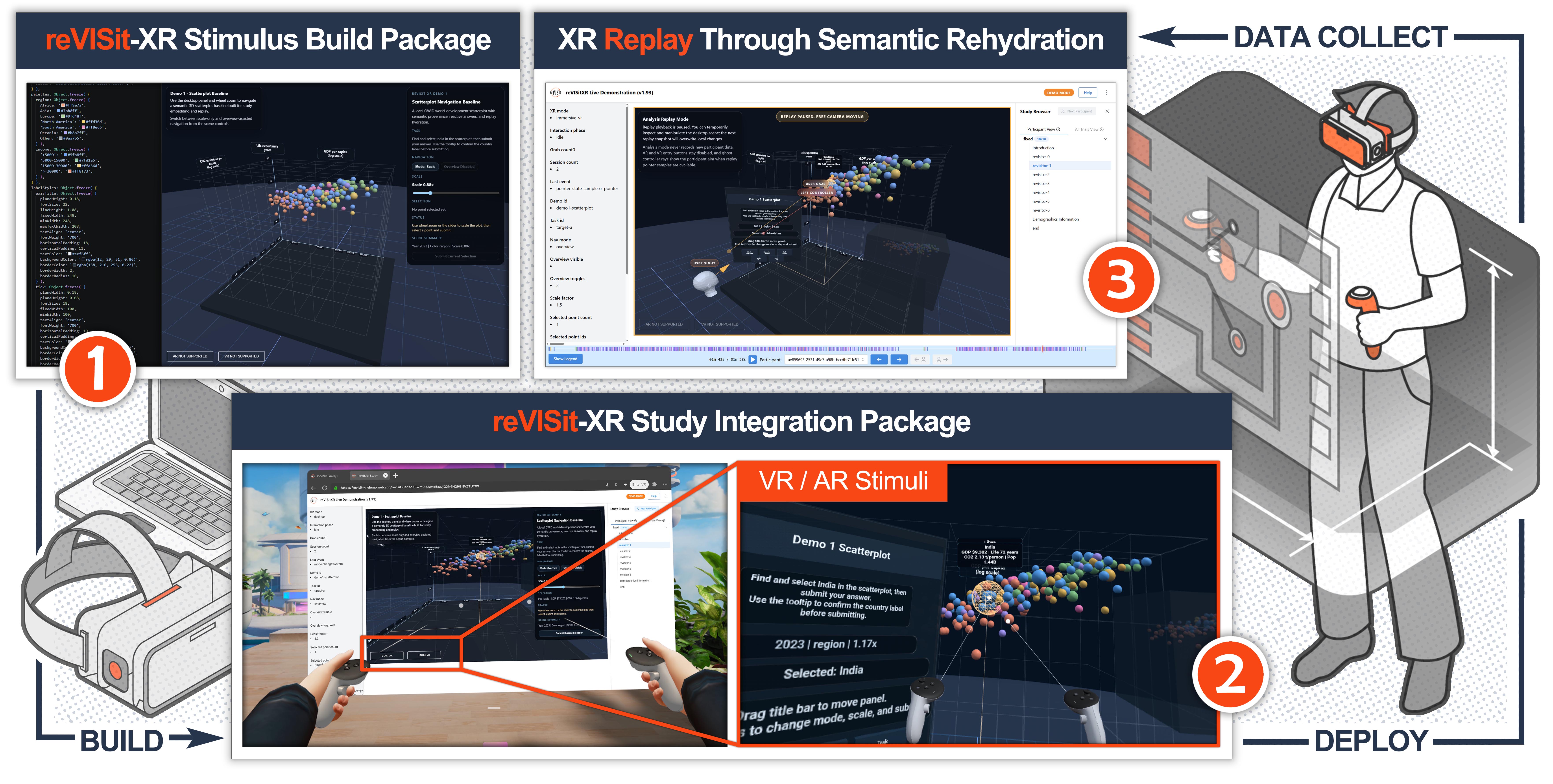}
  \caption{
  \textbf{\textit{reVISit-XR}} makes a WebXR stimulus embeddable, trackable, and replayable within a reVISit study. 
  \textbf{(1)} Researchers build or adapt a WebXR stimulus in the \textbf{reVISit-XR Stimulus Build Package}, 
  \textbf{(2)} embed it in a reVISit study through the \textbf{Study Integration Package} (or a study forked from \href{https://github.com/revisit-studies/study/}{the reVISit GitHub repository\(\nearrow\)} \cite{ding2023revisit,cutler2026revisit}), then collect participant responses and semantic provenance during headset interaction, 
  \textbf{(3)} and later inspect the session through semantic state rehydration.
  }
  \label{fig:teaser}
}

\abstract{
    Extended reality (XR) is increasingly a setting for empirical visualization research, where study-relevant state is distributed across headset and controller pose, scene configuration, selections, spatial layouts, and AR anchors. 
    Existing XR tools support parts of the workflow, such as scene authoring, interaction, or session analysis, yet seldom treat an XR stimulus as a reusable component of a complete study lifecycle. 
    We present \textbf{\textit{reVISit-XR}}, an extension of reVISit that makes customizable WebXR stimuli embeddable, trackable, and replayable within empirical visualization studies. 
    reVISit-XR sequences XR scenes alongside standard study components, collects reactive task responses, captures scene-authored semantic state together with generic XR traces, and rehydrates participant sessions for later desktop and headset analysis. 
    Organized as a reusable stimulus build package and a study integration package, it lets XR stimuli act as first-class study components. 
    We demonstrate its scope and feasibility through seven integrated, reusable examples and a deployed study, and discuss its current capabilities and future extensions.
} 

\keywords{Human-centered computing, extended reality, visualization systems and tools, research infrastructure}

\begin{document}


\firstsection{Introduction}

\maketitle


Extended reality (XR) has become an important setting for visualization research, both as an immersive display medium and as a site for studying embodied analytic behavior \cite{ens2022immersive,fonnet2019survey,kraus2021value}. 
XR studies ask how participants navigate abstract 3D data, compare geographic flows, arrange analytic views around themselves, anchor information to physical surfaces, interpret physical scale, or coordinate gaze, hand, and controller input during high-frequency interaction \cite{6dehesa2020touche,ens2022immersive,5lee2020data,3lee2020shared,4_2willett2016embedded,1yang2020embodied,2yang2018origin,4zhu2024make}. 
These settings raise a methodological challenge for empirical infrastructure: meaningful study state may be distributed across headset and controller pose, scene configuration, selected objects, spatial layouts, AR anchors, interaction events, and task responses. 
Recording too little leaves later interpretation ambiguous, while dense frame-by-frame traces are hard to inspect, replay, or connect to task-level behavior \cite{cutler2020trrack,ding2023revisit,lee2022remotelab}.


Consider a situated AR study in which a participant places a data overlay on a physical surface, changes metrics, selects an object, and later submits an answer. 
Beyond rendering the scene, the study must carry task-relevant states such as AR anchors and participant actions and answers across deployment, data collection, and analysis. 
Existing XR and WebXR libraries handle scenes rendering and interaction, while XR authoring and study tools support important parts of XR experimentation, including immersive visualization authoring, interaction prototyping, device/input management, task specification, and remote VR study execution \cite{butcher2020vria,cordeil2019iatk,gomes2020extended,gonzalez2024xdtk,lee2022remotelab,lobo2020flex}.
What is still missing is a full study-lifecycle integration in which an XR stimulus becomes a reusable empirical study component, sequenced with the rest of a study and inspectable after deployment. 
\textit{reVISit} \cite{cutler2026revisit,ding2023revisit,wilburn2025running} provides this lifecycle for conventional visualization studies, covering study specification, deployment, standard components such as consent and questionnaires, response collection, provenance, and downstream analysis, which makes it a natural foundation for bringing 
XR stimuli into empirical study practice. 


We present \emph{\textbf{reVISit-XR}}, an extension of the \href{https://revisit.dev/}{reVISit\(\nearrow\)} toolkit and framework \cite{cutler2026revisit,ding2023revisit,wilburn2025running} that makes customizable WebXR stimuli embeddable, trackable, and replayable within empirical visualization studies.
It lets researchers embed XR scenes as study components, collect task-facing responses from within those scenes, record scene-authored semantic state alongside generic XR traces, and rehydrate participant sessions for desktop and headset analysis. 
Here, semantic state refers to compact task-level variables 
to describe participant activity, distinct from raw controller streams or every rendered object transform \cite{heer2008graphical,ragan2015characterizing,xu2020survey}.
Given baseline reVISit does not include functionality for tracking the multiple streams of provenance and associated analysis capabilities needed in XR, our contribution is a study-infrastructure layer whose unit of support is the XR stimulus as first-class, reusable empirical study component: deployable within a reVISit study flow, captured as interpretable semantic provenance, and replayable through semantic-state rehydration of what a participant 
saw, selected, arranged, or triggered at a given moment. 

Following the demonstration method of prior reVISit work \cite{ding2023revisit}, we show the scope and feasibility of the infrastructure through seven integrated examples \cite{6dehesa2020touche,5lee2020data,3lee2020shared,1yang2020embodied,2yang2018origin,4zhu2024make}.
These examples serve as reusable templates, illustrating how diverse XR visualization cases can be deployed, instrumented, and replayed within a common empirical workflow.
In the spirit of an applications note, this short paper 
responds to renewed calls to support the increasing complexity of empirical visualization studies, including a recent ubiquitous-analytics agenda that names a spatial and immersive extension of reVISit-style evaluation infrastructure as an open challenge \cite{elmqvist2026somewhere}, while easing practical pain points for research teams.
We also discuss the need for more coordinated efforts to elicit technical needs and challenges in the XR space, to develop better formalizations for XR user studies which can help guide future efforts while improving reproducibility and transparency.

\section{Background}

Prior XR visualization work shows that immersive systems are studied as environments in which navigation, spatial organization, situated context, and interaction modality shape analytic behavior \cite{ens2022immersive,fonnet2019survey,kraus2021value}. 
Researchers have examined how users navigate abstract 3D data \cite{1yang2020embodied}, compare origin-destination flows on maps and globes \cite{2yang2018origin}, arrange multiple views in a surrounding workspace \cite{3lee2020shared}, read information anchored to real-world surfaces \cite{4zhu2024make}, interpret physical scale through embodied experience \cite{5lee2020data}, and coordinate gaze, hand, and controller input in immersive tasks \cite{6dehesa2020touche}. 

A parallel line of systems work has lowered the cost of building and studying immersive visualizations. 
IATK \cite{cordeil2019iatk} and DXR \cite{3sicat2018dxr} support authoring immersive visualizations and interaction techniques, while VRIA \cite{butcher2020vria} shows the value of browser-native immersive analytics deployment.
XDTK \cite{gonzalez2024xdtk} targets cross-device XR input and interaction.
Closer to study practice, RemoteLab \cite{lee2022remotelab} runs synchronized remote VR studies with multimodal recording and replay, and Flex-ER \cite{lobo2020flex} specifies immersive-analytics conditions, tasks, prototyping, debugging, and result sharing through a JSON-based environment. 
A further group of tools focuses on analyzing XR studies after the fact: MRAT \cite{nebeling2020MRAT} instruments mixed-reality applications through visual editors and reports task-success metrics for in-situ analysis of recorded sessions, and ReLive \cite{hubenschmid2022ReLive} couples an in-situ VR view with an ex-situ desktop view to revisit interactive spatial recordings of prior studies. 
reVISit-XR targets a different unit of support: the XR stimulus as a reusable empirical study component embedded in a broader reVISit workflow (reactive answers, semantic provenance, stored records, and replayable analysis).
Because it inherits this workflow, an XR study built with reVISit-XR composes with other parts of the reVISit suite, such as 
the reVISit \href{https://revisit.dev/docs/designing-studies/think-aloud/}{think-aloud\(\nearrow\)} technique \cite{cutler2025crowdsourced} or its \href{https://revisit.dev/docs/designing-studies/record-screen/}{video recording\(\nearrow\)}.
Methodologically, reVISit-XR builds on reVISit and Trrack.
Prior reVISit work \cite{cutler2026revisit,ding2023revisit,wilburn2025running} argues that visualization experiments benefit from reusable infrastructure for study specification, deployment, monitoring, stimuli integration, and downstream analysis. 
Trrack \cite{cutler2020trrack} and related provenance work \cite{heer2008graphical,ragan2015characterizing,xu2020survey} show that interaction histories are most useful when they preserve meaningful application state for recovery. 
In XR, however, 
logging every object transform is costly and difficult to interpret \cite{lee2022remotelab}.
reVISit-XR therefore adopts semantic-state rehydration, recording compact scene-level state alongside generic XR traces so that replay can restore task-relevant conditions.

\begin{figure*}[t]
  \centering
  \begin{subfigure}[b]{0.695\columnwidth}
  	\centering
  	\includegraphics[width=\textwidth]{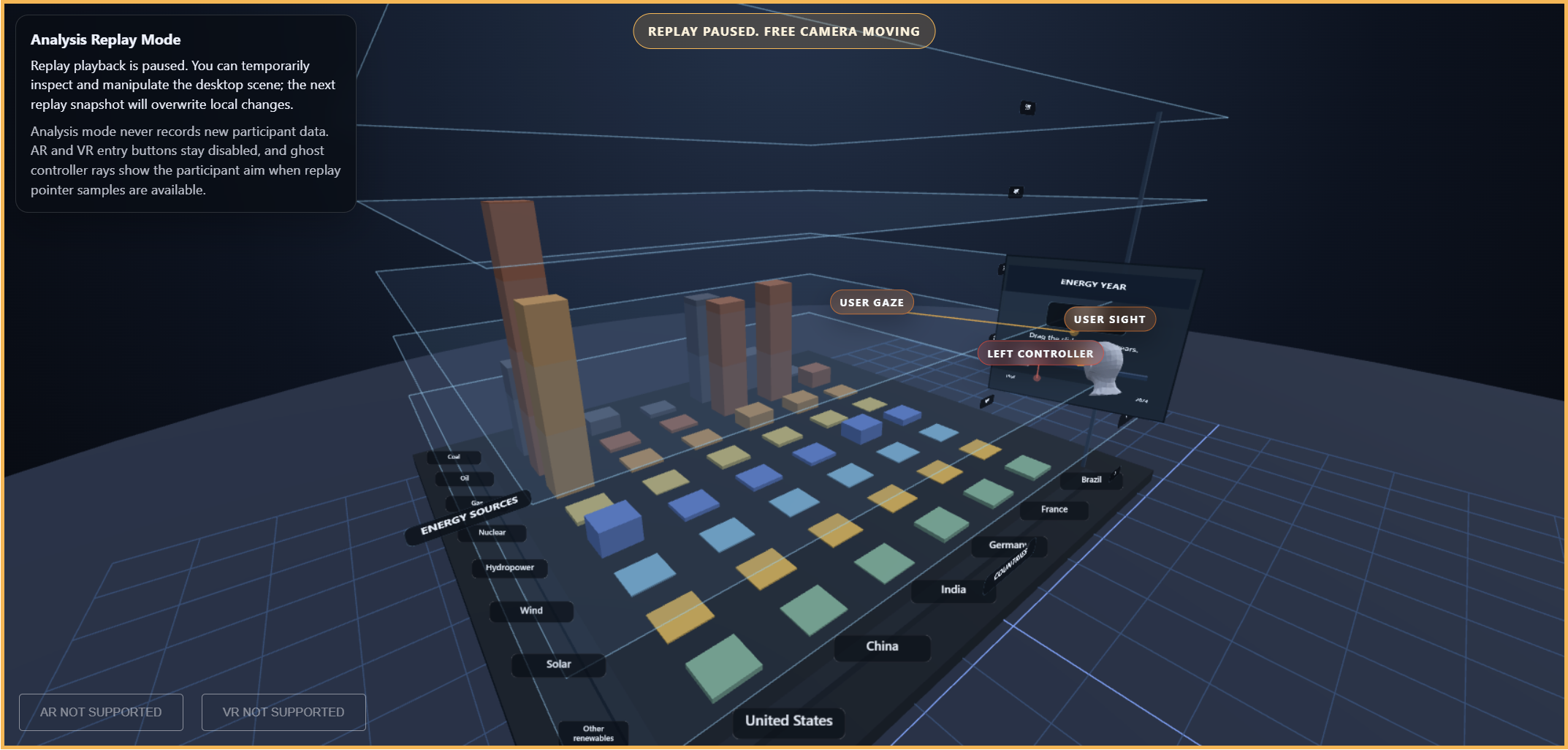}
  	\caption{Example: \textbf{\textit{revisitxr-0}}}
  	\label{fig:revisitxr-0}
  \end{subfigure}%
  \hfill%
  \begin{subfigure}[b]{0.695\columnwidth}
  	\centering
  	\includegraphics[width=\textwidth]{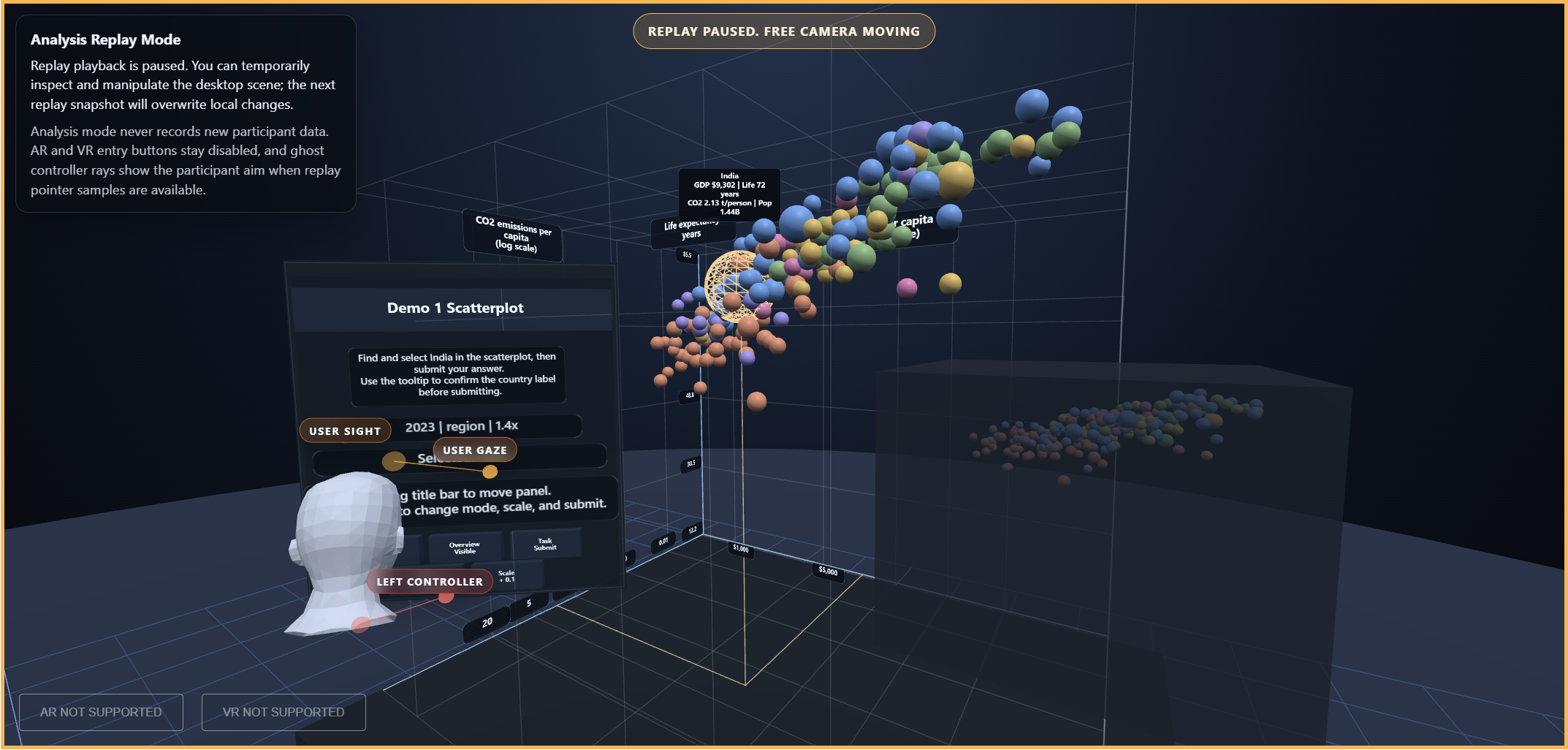}
  	\caption{Example: \textbf{\textit{revisitxr-1}}}
  	\label{fig:revisitxr-1}
  \end{subfigure}%
  \hfill%
  \begin{subfigure}[b]{0.695\columnwidth}
  	\centering
  	\includegraphics[width=\textwidth]{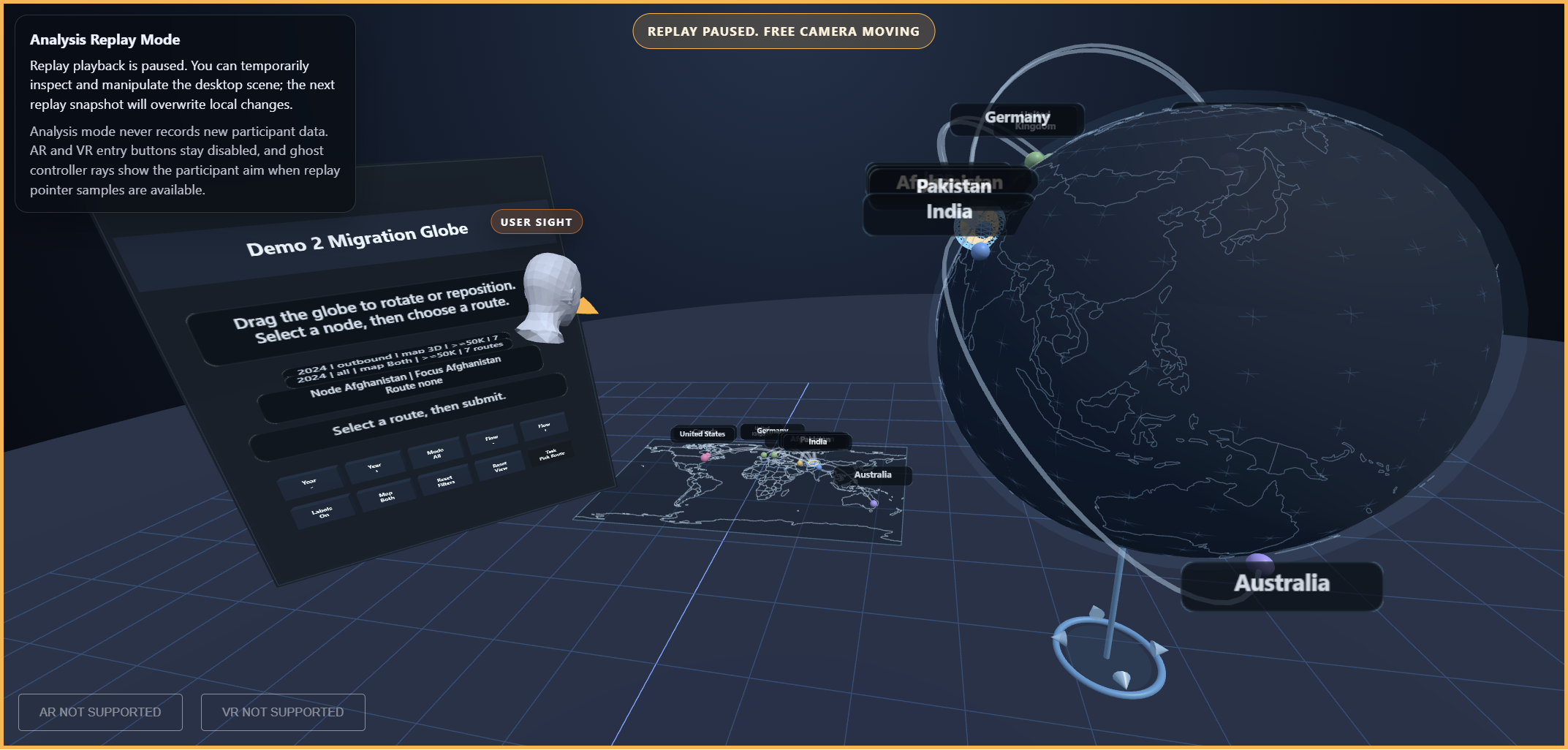}
  	\caption{Example: \textbf{\textit{revisitxr-2}}}
  	\label{fig:revisitxr-2}
  \end{subfigure}%
  \\%
\vspace{1pt}
  \begin{subfigure}[t]{0.52\columnwidth}
  	\centering
  	\includegraphics[width=\textwidth]{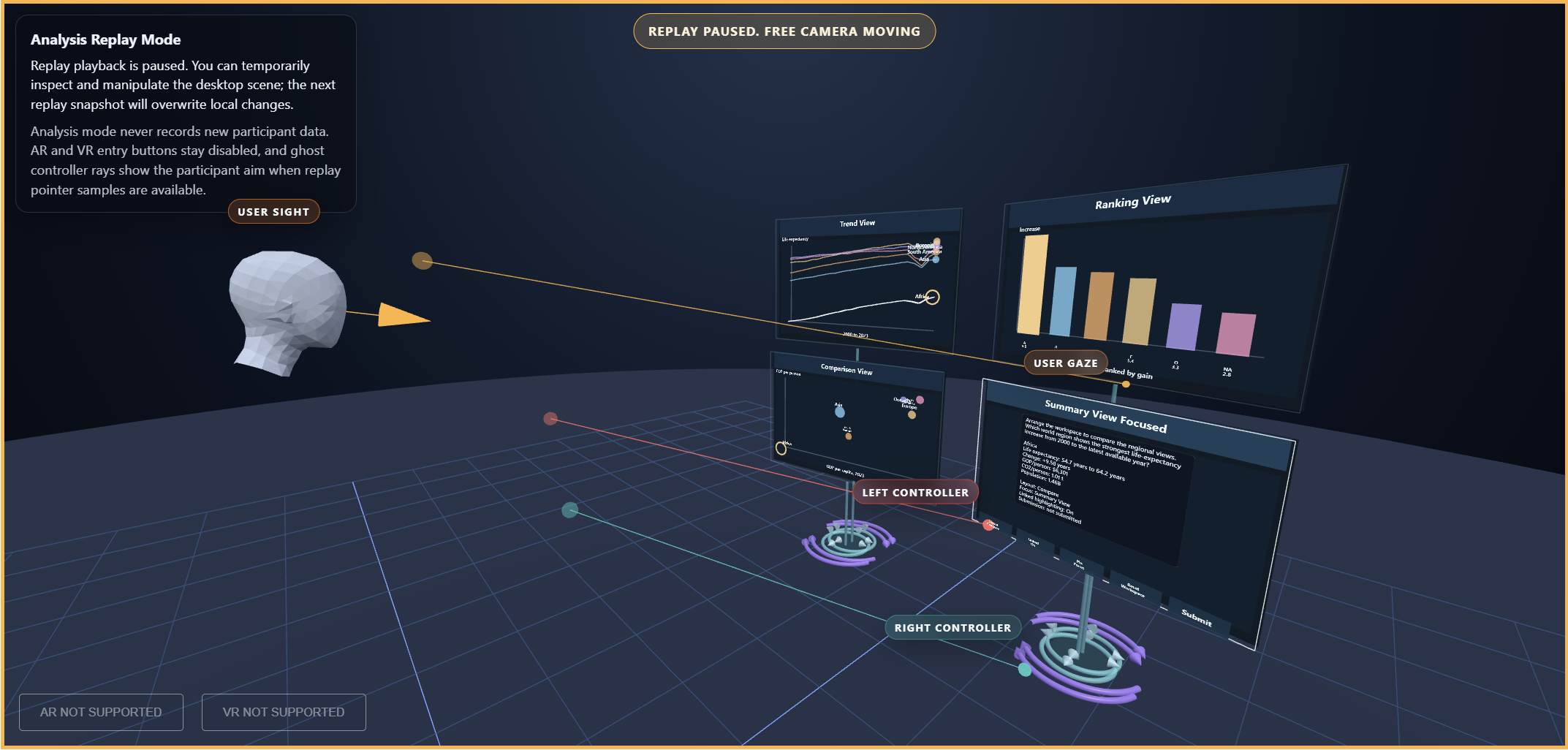}
  	\caption{Example: \textbf{\textit{revisitxr-3}}}
  	\label{fig:revisitxr-3}
  \end{subfigure}%
  \hfill%
  \begin{subfigure}[t]{0.52\columnwidth}
  	\centering
  	\includegraphics[width=\textwidth]{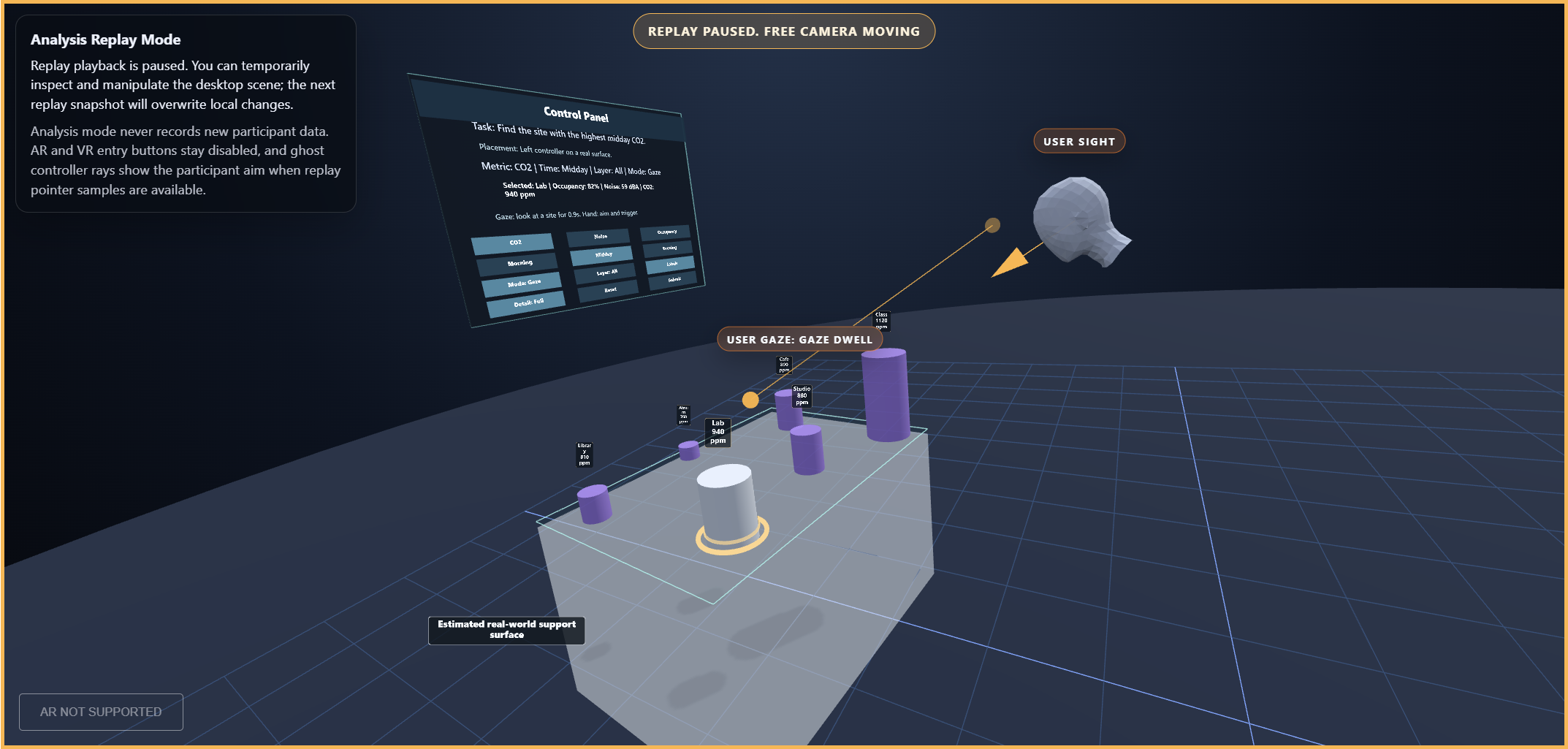}
  	\caption{Example: \textbf{\textit{revisitxr-4}}}
  	\label{fig:revisitxr-4}
  \end{subfigure}%
  \hfill%
  \begin{subfigure}[t]{0.52\columnwidth}
  	\centering
  	\includegraphics[width=\textwidth]{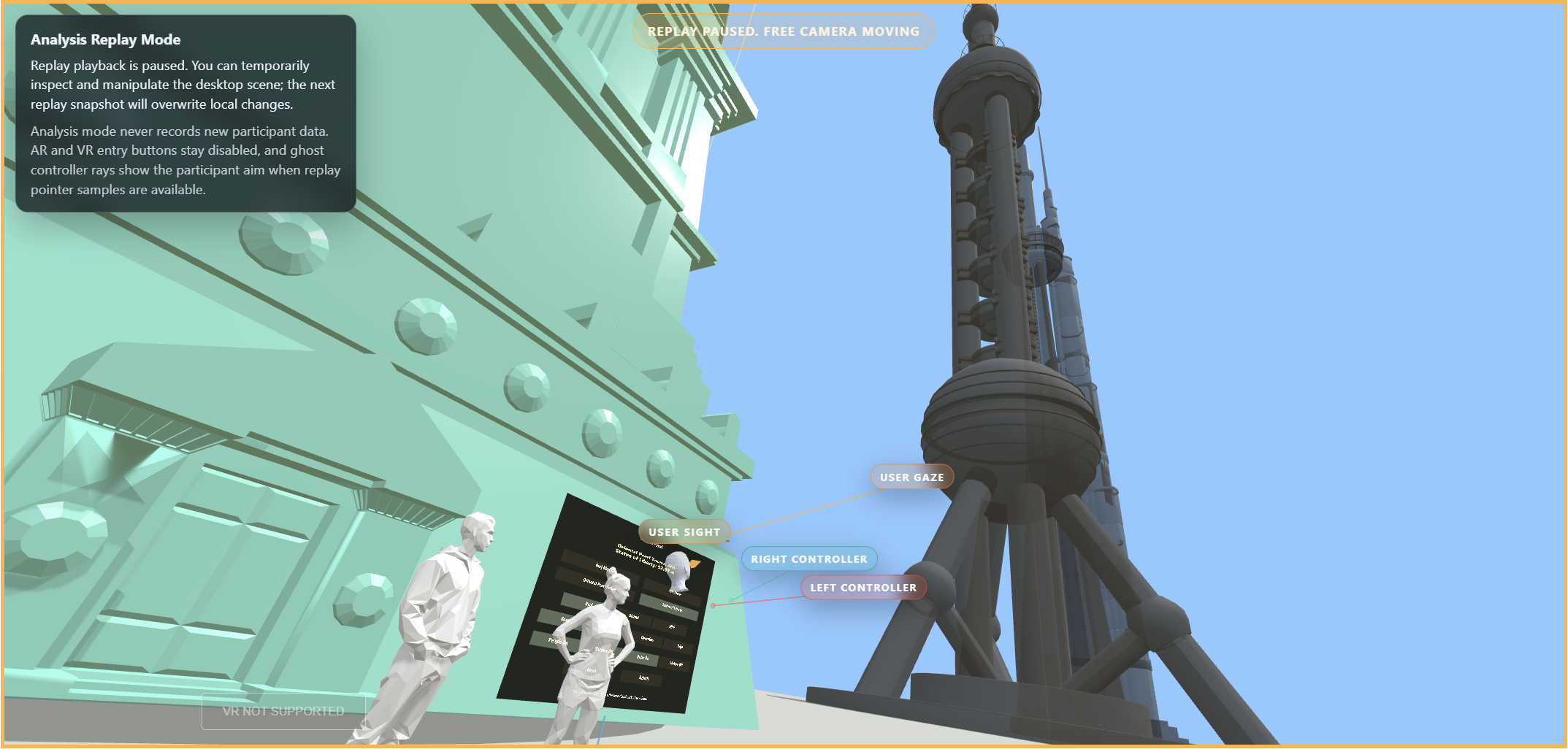}
  	\caption{Example: \textbf{\textit{revisitxr-5}}}
  	\label{fig:revisitxr-5}
  \end{subfigure}%
  \hfill%
  \begin{subfigure}[t]{0.52\columnwidth}
  	\centering
  	\includegraphics[width=\textwidth]{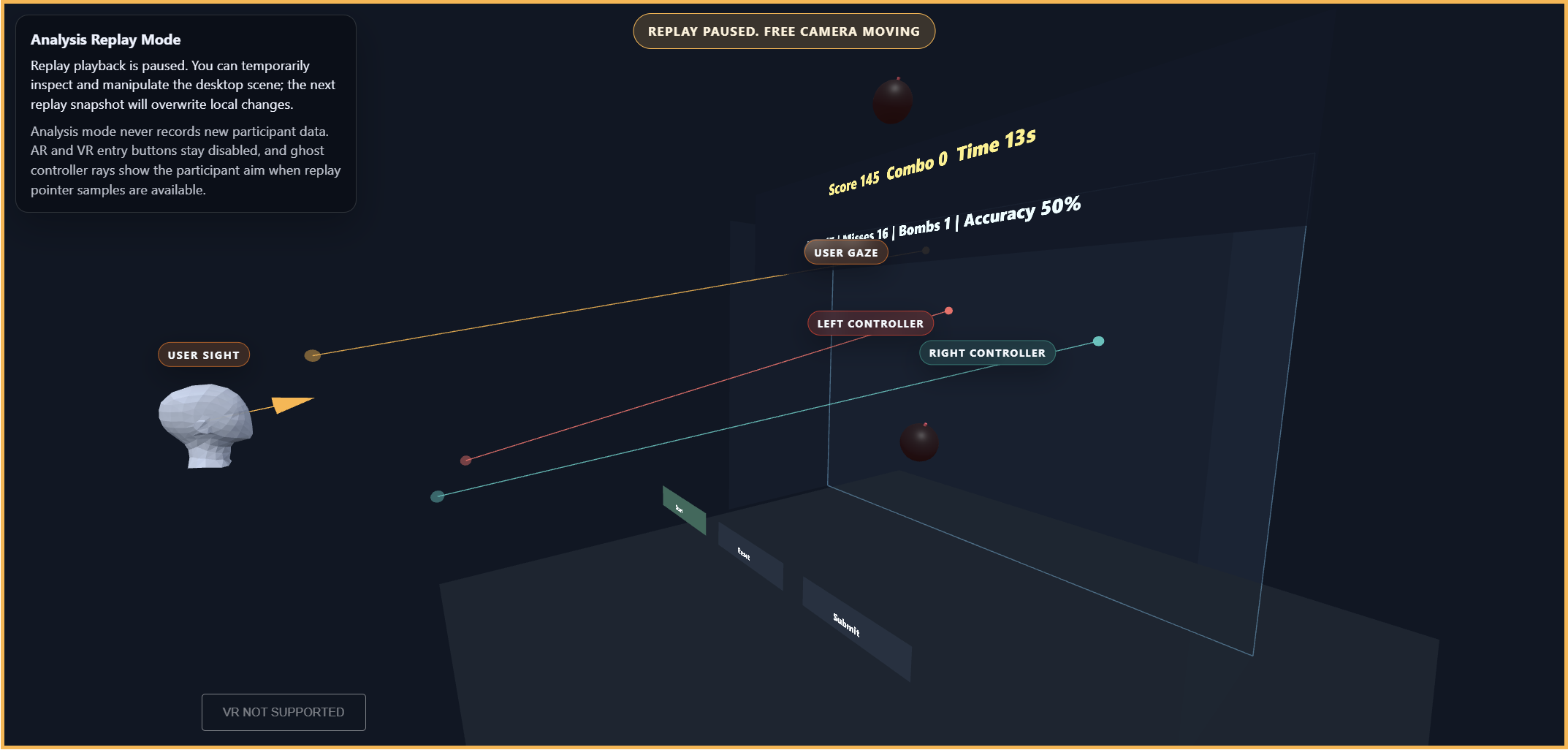}
  	\caption{Example: \textbf{\textit{revisitxr-6}}}
  	\label{fig:revisitxr-6}
  \end{subfigure}%
  \vspace{1pt}
  \subfigsCaption{
  Seven reusable example stimuli that cover recurring XR study-state requirements: 
  (\textbf{a, b}) immersive abstract visualization; 
  (\textbf{c}) geospatial immersive flow map; 
  (\textbf{d}) spatial layout in a multi-view workspace; 
  (\textbf{e}) anchoring situated AR overlay; 
  (\textbf{f}) scale comparison in landmark-scale; and 
  (\textbf{g}) compact game event logging.
  }
  \label{fig:examples}
  \vspace{-6pt}
\end{figure*}


\section{The reVISit-XR Framework}

reVISit-XR treats a WebXR stimulus as a first-class empirical study component, so an XR study can be deployed, instrumented, and later reconstructed within a single reVISit workflow. 
Rendering libraries already handle scenes and interaction; the difficulty lies in the study infrastructure around them, which in XR is non-trivial for \textbf{five} reasons. 
(i) \textit{\textbf{Design Requirement 1}} (marked as \textbf{DR1}): Study-relevant state is distributed across headset and controller pose, scene configuration, selections, spatial layouts, and AR anchors, so the system must distinguish task-relevant state from incidental rendering. 
(ii) \textbf{DR2}: The stimulus must act as a study component, sequenced with the rest of the study (instructions, consent, questionnaires) without forcing the participant to remove the headset. 
(iii) \textbf{DR3}: Tasks must expose answers the study can store and analyze without scene-specific coupling. 
(iv) \textbf{DR4}: Provenance must be captured without logging every frame, which is costly and hard to interpret in XR \cite{cutler2020trrack,lee2022remotelab}. 
(v) \textbf{DR5}: Live participation and analysis replay must stay distinct, and a session must be reconstructable after deployment. 
reVISit-XR meets these as a study infrastructure layer: 
It follows reVISit's architecture \cite{cutler2026revisit,ding2023revisit} 
for specification, deployment, and analysis, and adapts Trrack's provenance model \cite{cutler2020trrack} to XR state 
with compact runtime context like camera pose and replay-pointer state. 
The rest of this section describes how each requirement is met (\cref{fig:teaser}). 

The implementation is organized around two complementary packages: (i) the \href{https://github.com/revisit-games/reVISitXR-scene1}{\textbf{\textit{reVISit-XR Stimulus Build Package}}\(\nearrow\)}, and (ii) the \href{https://github.com/revisit-games/reVISitXR-studies}{\textbf{\textit{reVISit-XR Study Integration Package}}\(\nearrow\)}. 
The Stimulus Build Package is a reusable WebXR runtime containing the scene registry, shared Three.js/WebXR scaffolding, input handling, logging and serialization utilities, replay visuals, reusable scene helpers, and authored XR stimuli. 
The Study Integration Package embeds the built runtime as reVISit website components, defines study flow and component configuration, and forwards provenance, answers, and analysis-control messages to the embedded iframe.

\subsection{XR Scene Contract}

Each XR stimulus is implemented as a scene definition that declares a scene key, URL query value, display label, supported immersive modes, logging configuration, optional visual and behavioral configuration, state normalization logic, and a \verb|createScene| function. 
Here, \textbf{XR} spans both \textbf{VR} and \textbf{AR}: a scene declares which immersive modes it supports, so the same infrastructure can serve an immersive VR scene, a situated AR scene, or both, depending on the study (\cref{fig:examples}).
The central scene registry resolves the active scene from the URL and activates the corresponding scene controller. 
At runtime, each scene receives a context object that exposes shared services: access to the Three.js scene and camera, the current presentation mode, interaction policy, controller state, scene-state logging, replay hydration hooks, and answer synchronization. 
By owning their domain logic and state normalization, scenes decide which variables count as task-relevant (\textbf{DR1}),
while the shared runtime handles participant-state export, replay suppression, controller metadata, and parent-study communication. 

\subsection{Study Communication \& Reactive Answers}


XR stimuli are embedded through reVISit's existing website-component pathway, with reVISit-XR adding the XR-specific communication, provenance, analysis-control, and replay logic that headset studies need, so XR trials can be sequenced with ordinary components (\textbf{DR2}). 
Participants thus move through the study in the browser, enter embedded XR stimuli, and continue without repeatedly removing the headset.
Concretely, the two packages meet at an iframe boundary: the parent study is the reVISit page that drives the study flow, and the child runtime is the embedded reVISit-XR WebXR app it hosts. 
The child posts readiness, answer, and provenance messages to the parent and receives study data, prior answers, provenance snapshots, and analysis-control updates (\cref{fig:teaser}). 
Opened outside the study wrapper, the same runtime still works as a standalone WebXR stimulus for development and debugging. 
On the study side, the integration package builds iframe URLs, appends trial and iframe identifiers, and forwards an analysis-control object that distinguishes study mode from analysis mode, and paused replay from active playback (\textbf{DR5}).
Each scene then exposes task-specific fields through a \verb|getAnswerSummary| hook, such as selected data items, map filters, panel layouts, AR placement state, comparison mode, or game outcomes (\textbf{DR3}). 



\subsection{Semantic Provenance \& Interaction Policy}

The central logging principle in reVISit-XR is action- and task-level semantic state capture (\textbf{DR4}). 
A semantic state field is a scene-authored application-state variable \cite{heer2008graphical,ragan2015characterizing,xu2020survey} that a researcher would use to explain or replay what happened, for example the selected datum, the active layout or comparison mode, whether an AR anchor is confirmed, or a game's score and seed.  
Semantic snapshots are event-driven: a scene records a new state when an authored, study-relevant action occurs, such as selecting a data item or switching a condition. 
Generic XR traces, such as camera, controller, object-transform, and pointer samples, are handled separately through runtime sampling controls, with default settings that each scene can override.  
In practice, reVISit-XR captures committed, task-relevant changes together with sampled pose traces, and deliberately does not persist per-frame mesh transforms, transient hover or animation state, or raw engine objects, which are costly to store and seldom needed to reconstruct a task condition. 
These choices give reVISit-XR three layers of state: 
(i) shared runtime context, 
(ii) scene-specific semantic state, and 
(iii) compact reactive summaries exported to the parent study (\textbf{DR1}). 
This separation keeps the runtime, the study wrapper, and individual scenes loosely coupled. 
Interaction policy enforces the live/replay boundary (\textbf{DR5}): in analysis mode, immersive entry is disabled; while playback runs, replayed participant state drives the scene and local interaction is blocked; and when playback is paused, desktop inspection is allowed, but temporary edits are overwritten by the next replay snapshot and never create new participant data. 

\begin{table*}[t]
  \centering
  \caption{The seven reusable example stimuli in reVISit-XR (\cref{fig:examples}), with their input data, the XR stimulus and participant task, and the task-relevant semantic state that reVISit-XR records and rehydrates on replay. Modes follow each scene's declared immersive support.}
  \label{tab:usageexamples}
  \setlength{\tabcolsep}{3pt}
  \renewcommand{\arraystretch}{0.9}
  \setlength{\dashlinedash}{0.6pt}\setlength{\dashlinegap}{1.6pt}%
  {\scriptsize
  \begin{tabular}{@{}
    >{\columncolor[gray]{0.93}[2pt][2pt]\raggedright\arraybackslash}m{0.1\textwidth}
    >{\centering\arraybackslash}m{0.05\textwidth}
    >{\columncolor[gray]{0.93}[2pt][2pt]\raggedright\arraybackslash}m{0.2\textwidth}
    >{\raggedright\arraybackslash}m{0.285\textwidth}
    >{\columncolor[gray]{0.93}[2pt][2pt]\raggedright\arraybackslash}m{0.31\textwidth}@{}}
    
    \toprule
    
    \textbf{Example} & \textbf{Mode} & \textbf{Input Data} & \textbf{XR Stimulus \& Participant Task} & \textbf{Captured Semantic State} \\
    
    \midrule
    
    \textit{revisitxr-0}~\cite{1yang2020embodied} & VR/AR &
    OWID per-capita energy by source. &
    3D bar matrix; select bars and read values via an XR/desktop panel. &
    Selected bar, active filter, panel transform, answer. \\
    \hdashline
    \textit{revisitxr-1}~\cite{1yang2020embodied} & VR/AR &
    OWID country--year indicators (GDP, life expectancy, CO\textsubscript{2}, pop.). &
    Immersive 3D scatterplot; orbit/zoom, overview toggle, point selection; e.g., locate or compare countries. &
    Nav mode, overview state + toggles, scale, selected point IDs, answer. \\
    \hdashline
    \textit{revisitxr-2}~\cite{2yang2018origin} & VR/AR &
    OWID emigrant stock by destination (Afghanistan) and total immigrant stock. &
    Linked 3D globe + flat map; switch globe/map, year slider, stock threshold, select country or route. &
    Year, route direction, threshold, focused country, selected route, labels, map/globe mode and pose. \\
    \hdashline
    \textit{revisitxr-3}~\cite{3lee2020shared} & VR/AR &
    OWID region-level indicators. &
    Workspace of floating panels; drag/rotate panels, switch layout presets, focus, link highlights, select datum. &
    Layout mode, focused/selected view, selected datum, linking, visible/pinned views, panel transforms, answer. \\
    \hdashline
    \textit{revisitxr-4}\textsuperscript{\dag}~\cite{4zhu2024make,4_2willett2016embedded} & AR &
    Campus environmental readings (CO\textsubscript{2}, noise, occupancy) by site and time. &
    Overlay anchored to a real surface via WebXR hit-test; inspect site markers, toggle metric, select by gaze or controller. &
    Placement confirmed, anchor transform + surface height, metric/time/layer, labels, selected/focused site, modality. \\
    \hdashline
    \textit{revisitxr-5}~\cite{5lee2020data} & VR &
    Real-world heights of named landmarks with human-figure references. &
    To-scale VR landmarks, shadow/ruler cues, labels, and authored viewpoints; switch real-scale / distant / miniature comparison. &
    Selected landmark, comparison mode, viewpoint preset, cue toggles, answer. \\
    \hdashline
    \textit{revisitxr-6}~\cite{6dehesa2020touche} & VR &
    Procedurally generated targets from a deterministic seed. &
    \textit{Fruit Ninja}-style slicing~\cite{6_4_coutu2020immersiveness,6_3_islam2023emotional}; slice targets, avoid bombs with controller rays. &
    Round seed, discrete events, outcome stats; targets reconstructed from seed. \\
    \bottomrule
  \end{tabular}\par
  \vspace{1pt}
  \scriptsize\noindent\textsuperscript{\dag}\, In AR, reVISit-XR anchors the overlay to a detected real-world surface and records the anchor transform and surface height, but does not capture or reconstruct the surrounding real environment; analysis replay shows a virtual pedestal as a proxy for the real surface. 
  }
\vspace{-12pt}
\end{table*}

\subsection{Replay Through Semantic Rehydration}

Replay in reVISit-XR means reconstructing the participant-visible task condition at a chosen moment, and it works by semantic-state rehydration (\textbf{DR5}): rebuilding that condition from a compact, scene-authored state object instead of from a video or a full scene-graph dump. 
A snapshot pairs shared runtime context with each scene's compact \verb|sceneState| and active scene key. 
During analysis, the runtime normalizes the replayed state, activates the appropriate scene if necessary, reconstructs the visual condition for that timepoint, and draws analysis-only visuals such as ghost controller rays, gaze indicators, and the recorded user-sight avatar. 
This supports inspection that video or dense transform logging cannot easily provide: 
video fixes a single viewpoint and cannot be re-examined from a new camera position, while dense transform logs preserve motion but are large and difficult to interpret. 

\subsection{Deployment \& Data Management}

As an extension of reVISit, reVISit-XR provides both online (e.g., \href{https://firebase.google.com/}{Firebase\(\nearrow\)}) and offline storage pathways for XR studies. 
In both, it preserves XR records in structured formats suited to downstream analysis: 
nested JSON retains the semantic provenance needed for replay and rehydration, while exports support external analysis workflows. 
Deployment follows the standard reVISit workflow and separates XR stimulus authoring from study composition: the Stimulus Build Package's output is inserted into the Study Integration Package as a versioned asset, and the study is deployed through the same static-site pathway as reVISit studies \cite{cutler2026revisit,ding2023revisit}.

\section{Usage Examples}

To show the scope of reVISit-XR, we built seven reusable example stimuli (\cref{fig:examples}),
each adapted from a common XR visualization task in the literature \cite{6dehesa2020touche,5lee2020data,3lee2020shared,4_2willett2016embedded,1yang2020embodied,2yang2018origin,4zhu2024make} 
and each exercising a recurring study-state requirement. 
They are reusable study templates that 
researchers can clone and adapt.
The example stimuli were constructed from publicly accessible datasets obtained through \href{https://ourworldindata.org/}{\textbf{\textit{Our World in Data}}\(\nearrow\)} (OWID) \cite{OWID2025Energy,OWID2026GDP,OWID2025LifeExpectancy,OWID2025CO2,OWID2024Population,OWID2024MigrantStock,OWID2024AfghanEmigrants}.
\cref{tab:usageexamples} summarizes each example's input data, its XR stimulus and participant task, and the task-relevant semantic state that reVISit-XR records and rehydrates on replay.
All seven are integrated into the reVISit-XR Stimulus Build Package and into the deployed Study Integration Package as a \href{https://revisit-xr-demo.web.app/}{\textbf{live demo}\(\nearrow\)}. 
We also provide two \href{https://revisit-xr-demo.web.app/videos}{\textbf{short videos}\(\nearrow\)} that demonstrate testing and replaying the examples (the page may take 20–30s to load initially because of the video size).

\section{Discussion and Limitations}


reVISit-XR extends the broader reVISit ecosystem by making the XR stimulus a first-class, reusable study component, addressing the growing diversity of XR investigative interests while preserving compatibility with existing reVISit study workflows. 
We expect reVISit-XR to be most useful during the iterative move from pilots to a full study, where replaying inspectable sessions helps researchers catch and fix design problems while refining stimuli and tasks. 
Its reusable example also make it a practical resource for teaching XR and user-study design, letting students build, run, and inspect immersive studies without first assembling the surrounding infrastructure. 
Inheriting reVISit's open-source ethos, reVISit-XR is built to grow through community use, feedback, and extension 
by researchers working across different kinds of XR studies. 

reVISit-XR fits studies whose task-relevant condition can be expressed as compact semantic state, which covers the range our examples illustrate. 
Two design types sit at the edge of what it currently supports, both following from how it captures state. 
First, analyses that treat dense, frame-level motion as the primary record, rather than task-level semantic state, are bounded by logging cost:
in our testing, continuous interaction beyond roughly fifteen minutes in one uninterrupted scene introduced mild logging lag (consistent with Trrack \cite{cutler2020trrack}), so we recommend splitting long XR tasks into multiple trials or shorter scene segments.
Second, synchronous multi-user studies, such as the collaborative co-located sensemaking of Lee et al. \cite{3lee2020shared}, require several participants to share one environment in real time; 
this multi-user orchestration is not yet available in either reVISit or reVISit-XR, 
and we see it as a worthwhile future extension of the broader reVISit ecosystem.
Situated AR support is also bounded in scope: reVISit-XR captures the anchor and its surface but not the participant's real environment, so situated replays use a proxy surface; bringing real-world geometry into capture and replay (e.g., via passthrough) is future work.

\section{Future Work}

As with the broader reVISit ecosystem, we see reVISit-XR as one step in a longer infrastructure effort rather than a finished system. 
An immediate direction is community-informed expansion and evaluation. 
Recent reVISit work \cite{cutler2026revisit} shows the value of learning from researchers' real study practices through interviews and reflective iteration, and we plan to engage XR researchers in the same way. 
This is also how we intend to test reVISit-XR's central claim, that it makes XR studies easier to run, inspect, and reproduce: we plan to evaluate the toolkit through real studies built with it, observing in particular whether replay-based inspection helps researchers diagnose and improve their designs during the pilot-to-study iteration noted above.

\section{Conclusion}

We introduced reVISit-XR, an extension of reVISit that makes WebXR stimuli embeddable, trackable, and replayable as first-class study components. 
By combining stimulus building and integration, semantic provenance capture, and replay through state rehydration, it makes XR studies easier to deploy, inspect, and reproduce. 
We see reVISit-XR as one part of a broader effort toward open research infrastructure. 

\section*{Acknowledgment}

This work was supported in part by the National Science Foundation \#2213757. 
During the development of reVISit-XR, \textit{OpenAI Codex} (in \textit{Visual Studio Code}) was used only for debugging; 
No artificial intelligence (AI), large language models (LLMs), or similar systems were used to generate the design, prototyping, research, analysis, text, figures, or references in this work.

\section*{Supplemental Materials}
\label{sec:supplemental_materials}

The reVISit-XR Stimulus Build Package can be found at \href{https://github.com/revisit-games/reVISitXR-scene1}{github.com/revisit-games/reVISitXR-scene1\(\nearrow\)}. The reVISit-XR Study Integration Package can be found at \href{https://github.com/revisit-games/reVISitXR-studies}{github.com/revisit-games/reVISitXR-studies\(\nearrow\)}.
The live demo of the reVISit-XR study can be found at \href{https://revisit-xr-demo.web.app/}{revisit-xr-demo.web.app/\(\nearrow\)}.
The demonstration videos can be found at \href{https://revisit-xr-demo.web.app/videos}{revisit-xr-demo.web.app/videos\(\nearrow\)}.




\bibliographystyle{abbrv-doi-narrow}

\bibliography{REFS}
\end{document}